\documentclass[aps,pra,reprint,groupedaddress]{revtex4-1}

\usepackage{amsmath}
\usepackage{color}
\usepackage{amsfonts}
\usepackage{graphicx}
\usepackage{soul}
\usepackage[normalem]{ulem}

\usepackage{graphicx}
\usepackage{float}
\usepackage{url}

\usepackage[utf8]{inputenc}

\usepackage{amsmath,amssymb,amsfonts}
\usepackage{xcolor}

\newcommand{\op}[1]{\hat{#1}}
\newcommand{\definedas}{\equiv}

\begin{document}


\title{Superconducting circuit quantum computing with nanomechanical resonators as storage}


\author{Marek Pechal}
\email[]{mpechal@stanford.edu}

\author{Patricio Arrangoiz-Arriola}
\email[]{parrango@stanford.edu}

\author{Amir H. Safavi-Naeini}
\email[]{safavi@stanford.edu}

\affiliation{Department of Applied Physics and Ginzton Laboratory, Stanford University \\
 				348 Via Pueblo Mall, Stanford CA 94305 USA}


\date{\today}

\begin{abstract}
We analyze the quantum information processing capability of a superconducting transmon circuit used to mediate interactions between quantum information stored in a collection of phononic crystal cavity resonators. Having only a single processing element to be controlled externally makes this approach significantly less hardware-intensive than traditional architectures with individual control of each qubit. Moreover, when compared with the commonly considered alternative approach using coplanar waveguide or 3d cavity microwave resonators for storage, the nanomechanical resonators offer both very long lifetime and small size -- two conflicting requirements for microwave resonators. A detailed gate error analysis leads to an optimal value for the qubit-resonator coupling rate as a function of the number of mechanical resonators in the system. For a given set of system parameters, a specific amount of coupling and number of resonators is found to optimize the quantum volume, an approximate measure for the computational capacity of a system. We see this volume is higher in the proposed hybrid nanomechanical architecture than in the competing on-chip electromagnetic approach. 
\end{abstract}


\maketitle


\section{Introduction}\label{sec:intro}
Superconducting circuits are one of the architectures currently used to build the first coherent quantum devices with tens of quantum bits \cite{Devoret2013,girvin2011circuit,Kelly2018APS,IBMQExperience,Otterbach2017,xiang2013hybrid}, complex enough to preclude their efficient classical simulation. This exciting crossover to the regime where quantum devices may offer advantages in physical simulations or information processing over classical computers \cite{Preskill2018}, was enabled by rapid technological progress in the past decade, aimed mainly at the development of quantum gates with higher fidelities \cite{Barends2014} and qubits with longer coherence times \cite{Rigetti2012}. 

The prevailing approach to quantum computing with superconducting circuits is to use qubits as both data storage and processing units and to control each qubit individually. The second point, in particular, complicates scaling to large devices -- as the number of qubits grows, the amount of cabling and electronic equipment needed makes individual control of qubits challenging. Alternative approaches have emerged where instead of using the nonlinear element as a qubit, it is used as a processing element to control states in the larger Hilbert space of one~\cite{Ofek2016} or several~\cite{Naik2017} electromagnetic oscillators by application of more complex control signals. In the approach followed by Ref.~\cite{Naik2017} a system composed of a transmon qubit coupled to $N=11$ on-chip linear electromagnetic resonators is used effectively as an $N$-qubit system. Although the use of only a single or a small number of processing qubits presents a bottleneck in the computation and makes the process less parallelizable, control signals only need to be sent to the processing qubits, potentially saving a significant amount of resources. Such architectures are appealing since they can effectively amplify the quantum computational capacity of a physical setup.

Two-qubit gates are executed in series via the processing qubit in this architecture. Computation run times are therefore expected to be generally longer, and it is essential that the storage elements have very long coherence times to avoid excess loss in fidelity. When using microwave systems for storage, one can either use on-chip resonators \cite{Naik2017} (or qubits) or machined ``three-dimensional'' cavities \cite{Brecht2016}. On-chip resonators are usually compact but have coherence times on the same order as qubits, while 3d cavities can have orders of magnitude higher quality factors \cite{Reagor2016} but are challenging to scale due to incompatibility with microprocessing technologies.

Based on recent developments of hybrid systems combining superconducting circuits with mechanical resonators \cite{OConnell2010,Pirkkalainen2013,Chu2017,Satzinger2018,ArrangoizArriola2018,Noguchi2017,Manenti2017}, we propose an architecture in which on-chip mechanical resonators could serve as both very compact and long-lived quantum storage. In contrast with other similar electromechanical systems studied for example in \cite{Cleland2004} and \cite{McGee2013}, the method proposed here uses phononic crystal resonators which are micrometer-sized, and their quality factors can exceed $10^{10}$ \cite{MacCabe2018APS}. The use of phononic bandgap structures leads to robust high-$Q$ mechanical resonances. Moreover, phononic bandgaps isolate the qubit from phonon leakage channels~\cite{Ioffe2004} that are likely to become problematic on highly piezoelectric substrates such as those needed to obtain large coupling rates. Crucially, the small size of the resonators means that a substantial number of them can be fabricated in a space comparable with the size of a single qubit~\cite{ArrangoizArriola2018} and directly coupled to it. To make the resonators individually addressable by the qubit, they can be fabricated with sufficiently separated frequencies which are determined by the designed geometry of the phononic crystal sites, as described in section~\ref{sec:coupling}.

\section{Gate error analysis}\label{sec:gateError}
\begin{figure*}
\begin{center}
\includegraphics[width=17.5cm]{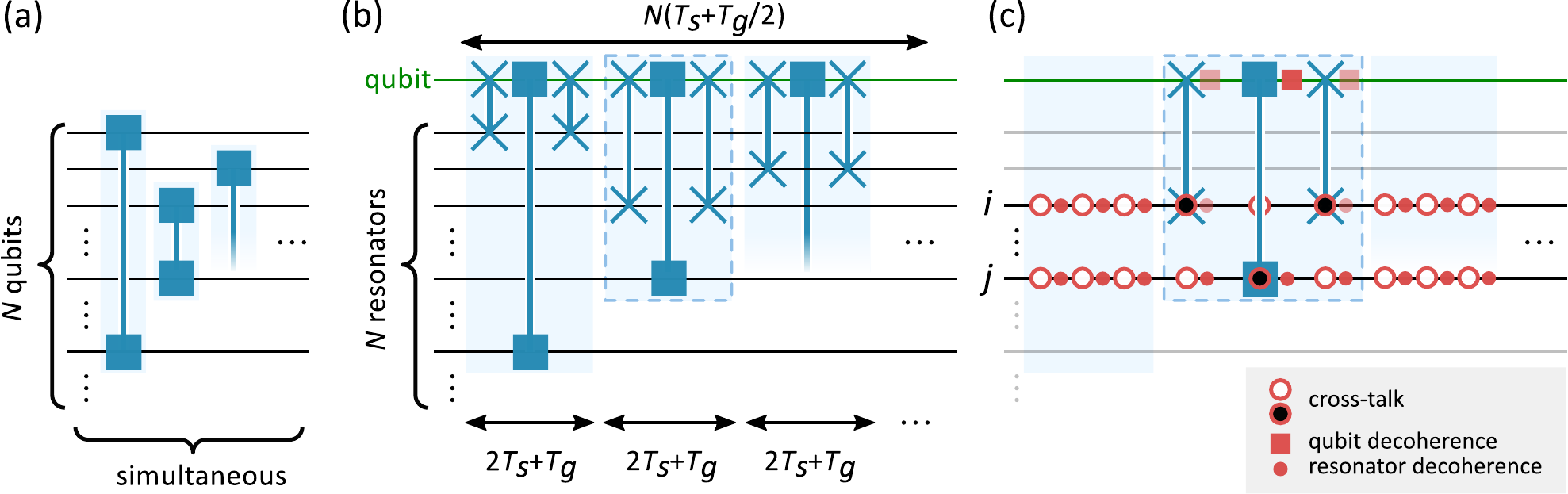}
\end{center}
\caption{(a) Schematic representation of one step of a quantum circuit acting on $N$ qubits. All two-qubit gates between distinct pairs of qubits are performed simultaneously. (b) Sequential version of the circuit from (a), with the single qubit mediating interactions between $N$ resonators. Each effective two-resonator gate consists of two qubit-resonator swap gates surrounding one arbitrary qubit-resonator gate. (c) The errors acting on a specific pair of resonators $i$ and $j$ in the sequential protocol. Decoherence errors are shown by solid circles (for the resonators) and squares (for the qubit). Cross-talk errors which occur because the gates acting on $i$ and $j$ also weakly address other resonators are shown by filled red circles. The other class of cross-talk errors, caused by gates performed on other resonators affecting $i$ and $j$, is indicated by empty red circles.}
\label{fig:errorAnalysis}
\end{figure*}

In the circuit picture of quantum computation, the algorithm is typically decomposed into a series of two-qubit and single-qubit gates. We will assume the single-qubit operations to be lumped into the two-qubit ones. Gates which operate on distinct pairs of qubits are assumed to be performed simultaneously in a single discrete time step. The number of such steps required to complete the computation is called the \emph{circuit depth}. An example of a single step in a circuit with $N$ qubits is shown schematically in Fig.~\ref{fig:errorAnalysis}(a). There can be up to $N/2$ two-qubit gates performed simultaneously between arbitrarily chosen pairs of qubits. 

In the architecture described here, the qubit states are stored in the resonators as superpositions of the vacuum state $|0\rangle$ and the single-photon Fock state $|1\rangle$. The gates, designed in such a way that the resonators do not leave this two-dimensional subspace, need to be performed sequentially via the single processing qubit.
The sequential equivalent of the circuit from Fig.~\ref{fig:errorAnalysis}(a) is illustrated in Fig.~\ref{fig:errorAnalysis}(b). A simple way to realize a gate between resonators $i$ and $j$ is to perform a swap operation between the qubit and resonator $i$, followed by an entangling gate between the qubit and resonator $j$ and another swap with $i$. Assuming that the swap operation takes a time $T_s$ and the entangling gate on average $T_g$, the effective gate between the resonators takes $2T_s+T_g$ and the whole step of the quantum circuit $N (T_s+T_g/2)$. For simplicity, we will consider the entangling gate to be a phase gate implemented as a $2\pi$ rotation swapping the resonator excitation into the qubit and back while accumulating an overall $-1$ phase factor \cite{Naik2017}. In this special case, we have $T_g = 2T_s$.

We will now focus on the evolution of one specific pair of resonators over the time period $N (T_s+T_g/2)$ and approximate it as an ideal two-qubit gate combined with an "error" acting on each of the resonators, occurring with some probability $\varepsilon$ which we would like to estimate. To this end, we will make use of some rather crude approximations, but we believe this does not greatly affect our main goal which is to observe how the performance of the system depends on the lifetimes of its components and how it scales with the number of resonators. For instance, we will not specify the precise nature of the errors (dephasing, relaxation, etc.) and will characterize them by a single "error probability". We will also assume that the error probabilities can be simply added together.

The evolution of two of the resonators, $i$ and $j$, is schematically illustrated in Fig.~\ref{fig:errorAnalysis}(c). The relevant error contributions, which are indicated by the red symbols, are due to the limited selectivity of the operations (cross-talk, which leads to some small amount of entanglement with resonators other than $i$ and $j$ or to excitation of the resonators to higher energy levels), and decoherence of the resonators and the qubit. As we will see, both types of errors present a trade-off between the number of resonators in the system and its performance. With increasing number of resonators, the time required for a single layer of the quantum circuit grows and the probability of a decoherence error goes up. At the same time, a larger number of resonators means a smaller detuning between them, leading to larger cross-talk errors.

To estimate the contribution to the error due to decoherence, we note that the quantum information stored in the resonator that we swap with the qubit spends roughly a time $T_s/2+T_g+T_s/2 = 3T_s$ in the qubit out of the total time $N(T_s+T_g/2)=2NT_s$. The corresponding error probability is therefore approximately $((2N-3)\Gamma_{r}+3\Gamma_{q})T_s$. The other resonator experiences the error rate $\Gamma_{r}$ for the whole period $2NT_s$. In total, the decoherence error probability \emph{per qubit} is
\[
  \varepsilon_{\mathrm{dec}} = 
  \left(
  \left(2N-\frac{3}{2}\right)\Gamma_{r}+\frac{3}{2}\Gamma_{q}
  \right)T_s \approx
  (
  2N\Gamma_{r}+3\Gamma_{q}/2
  )T_s.
\]
Here we have assumed $N\gg 1$ to simplify the expression.

To estimate the cross-talk error, we note that the ideal qubit-resonator gates considered above are resonant processes that are rotations in the subspace spanned by $|g1\rangle$ and $|e0\rangle$. The rate of this rotation is $2g$, where $g$ is the effective coupling strength between the qubit and the resonator. Assuming that the resonators are spaced uniformly in frequency space with a nearest-neighbor detuning $\delta$ and that they have the same coupling $g$ to the qubit, each of the gates drives unwanted transitions detuned by $\delta_k = \pm\delta,\pm 2\delta,\ldots$. In the limit of small $g/\delta$, we estimate the probability of these unwanted transitions as $\sum_k g^2/\delta_k^2 \propto g^2/\delta^2$. Numerical simulations indicate that this is a rather pessimistic estimate and by modulating the coupling $g$ smoothly in time, the cross-talk can be made significantly smaller. We will be conservative in our analysis and assume that the combined cross-talk error probability for the effective resonator-resonator gate, illustrated in Fig.~\ref{fig:errorAnalysis}(c) by the filled red circles, is $g^2/\delta^2$.

Cross-talk from gates between other pairs of resonators affects resonators $i$ and $j$, even if $i$ and $j$ are idle. We depict this schematically in Fig.~\ref{fig:errorAnalysis}(c) by the empty red circles. We may expect a gate acting on a resonator detuned by $\delta_k$ to induce an error with a probability at most about $g^2/\delta_k^2$. If we sum this expression over all the resonators we get a total which again scales as $g^2/\delta^2$. Once more, this is close to the worst-case estimate, and we expect to find a much more favorable dependence on $g/\delta$ by designing the coupling pulses with care.

Conservatively, all the cross-talk errors add up to an amount on the order of $g^2/\delta^2$. Importantly, this error probability does not explicitly scale with $N$. To keep this derivation brief, we did not discuss the potential constant pre-factor. We will denote it by $A$ and assume it is on the order of unity. Later we will see that the performance of the system depends only quite weakly on its exact value.

We consider the storage resonators frequencies to be uniformly distributed over the band gap of the phononic crystal. In silicon, gaps with frequency spans greater than half of their center frequency $\omega_0$ have been demonstrated~\cite{Alegre2010}. The nearest-neighbor detuning between the resonators is then $\omega_0/2N$. For numerical calculations, we will assume that $\omega_0/2\pi = 4\,\mathrm{GHz}$ which is compatible with typical superconducting qubit frequencies.

Since we assume that the qubit used in this system is a transmon -- a weakly anharmonic circuit -- we also need to take into account the presence of the transitions to its higher excited states. To first approximation, we consider only one spurious transition from the first to the second excited state which is detuned by $\alpha$ from the qubit's fundamental transition. With an appropriate choice of $\alpha$, we can ensure that whenever the qubit is effectively resonant with one of the resonators, the spurious transition frequency lies half-way between resonator frequencies and so is off-resonant by $\omega_0/4N$. We will therefore set $\delta$ equal to this smallest detuning encountered in the system.

Adding the cross-talk $A g^2/\delta^2 = 16 A N^2 g^2 / \omega_0^2$ and the decoherence contribution $\varepsilon_{\mathrm{dec}}$ together, we get the overall error probability $\varepsilon$ per qubit for a single step of the quantum circuit. We further note that the swap time $T_s$ is related to the coupling rate $g$ by $T_s = \pi/2g$. Now we can write the error probability $\varepsilon$ in a way that explicitly spells out its dependence on the number of resonators $N$ and on the coupling $g$:
\begin{equation}\label{eq:errorProb}
  \varepsilon(N) = 
  \frac{\pi (N\Gamma_{r} + 3\Gamma_{q}/4)}{g} + 
  \frac{16A N^2 g^2}{\omega_0^2}.
\end{equation}
As we will see below, {due to the trade-off between the cross-talk and decoherence contributions}, for a given set of decoherence parameters and number of resonators $N$, the error probability per qubit $\varepsilon$ is minimized for an optimal value of the coupling rate $g$.

\section{Quantum volume estimates}
We will now quantify the expected performance of the proposed electromechanical architecture in terms of the \emph{quantum volume} and show a favorable comparison with analogous systems using microwave resonators for storage \cite{Naik2017}. The quantum volume is a recently introduced figure of merit for quantum hardware \cite{Bishop2017} which captures the number of qubits in a system as well as the number of gates which can be performed with it, representing the intuitive notion that ``interesting" algorithms require both. If a system of a given type with $N$ qubits can implement ``typical" quantum circuits with maximum depth $d(N)$ before the error exceeds some fixed threshold, the quantum volume $V_Q$ is defined as 
\begin{equation}\label{eq:VQ}
  V_Q \definedas \max_{N}\left[\min(N,d(N))\right]^{2}
\end{equation}
The maximum depth $d(N)$ can be estimated as $1/N\varepsilon(N)$, where $\varepsilon(N)$ is the error probability per qubit in one step of the quantum circuit. This probability depends in a non-trivial way on the number of qubits due to various technical issues such as cross-talk, frequency crowding, etc. It is also strongly dependent on the topology of the system. For example, if the system has all-to-all connectivity between qubits then all two-qubit gates have in principle the same complexity. At the other extreme, if only nearest-neighbor couplings are available in a 1d chain of qubits then a typical two-qubit gate needs to be mediated on average by $N/3$ qubits and may therefore be expected to fail with a probability which grows linearly with $N$.

\begin{figure*}
\begin{center}
\includegraphics[width=16.5cm]{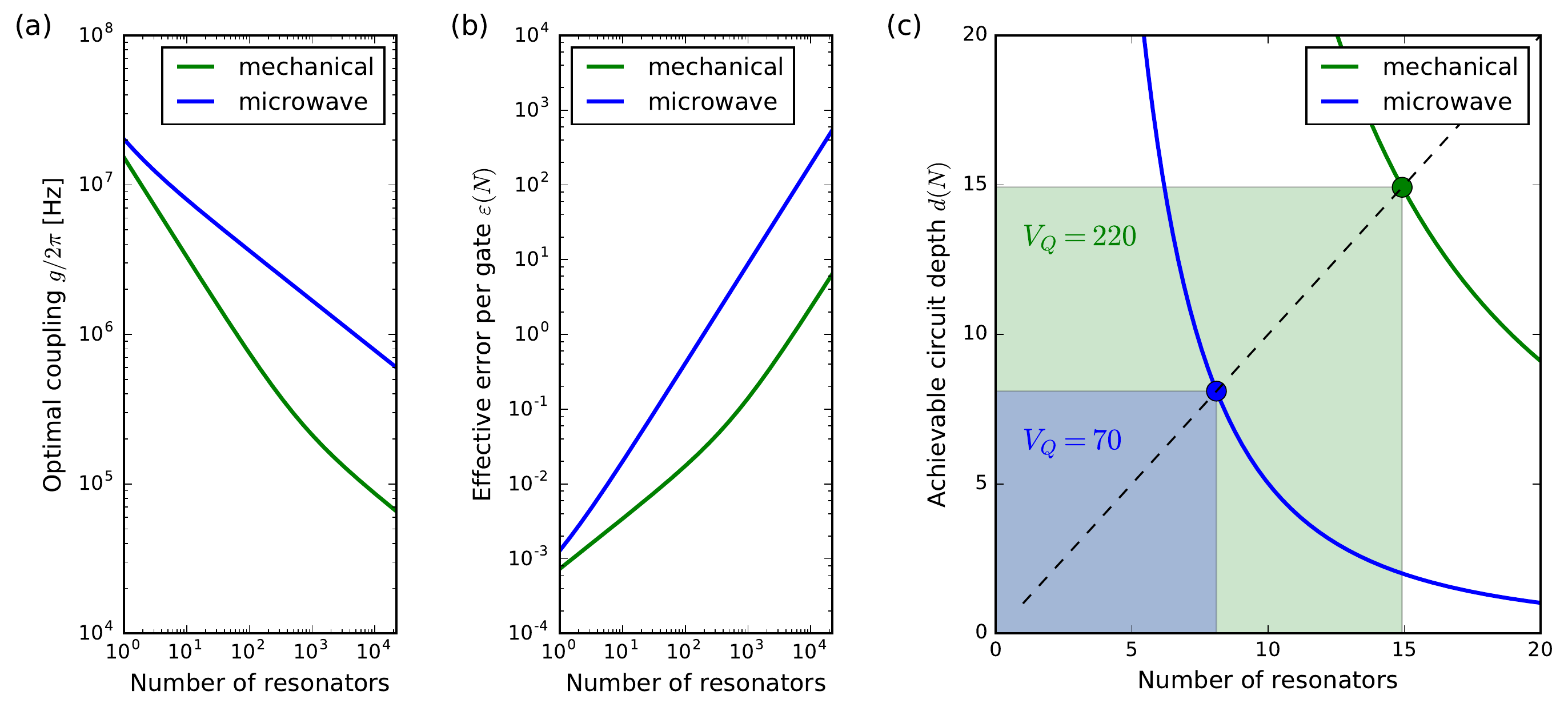}
\end{center}
\caption{(a) Dependence of the optimal coupling which minimizes the effective gate error on the number of resonators $N$. (b) The effective error probability per gate achieved at the optimal coupling from (a) as a function of $N$. (c) Illustration of the quantum volume in the electromechanical and purely microwave implementation in a plot of the achievable circuit depth $d(N)$ as a function of $N$. Since $d(N)$ decreases with increasing $N$, the quantity $\min(N,d(N))$ is maximized when $N=d(N)$, as shown by the points indicating the intersection of the curves with the dashed diagonal line. The quantum volume is then the area of the filled squares.}
\label{fig:VQplots}
\end{figure*}

In our system, the error probability is estimated by Eq.~(\ref{eq:errorProb}).
For any given $N$ and decoherence rates $\Gamma_{q}$ and $\Gamma_{r}$, we should choose $g$ to minimize this expression. The minimum is attained for 
\begin{equation}\label{eq:optg}
  g = \left(\frac{
  \pi (N\Gamma_r+3\Gamma_q/4)\omega_0^2
  }{
  32AN^2
  }\right)^{1/3}
\end{equation}
and takes the value
\begin{equation}\label{eq:optError}
  \varepsilon(N) \approx 
  3
  \left(\frac{
  4A \pi^2 N^2(N\Gamma_r+3\Gamma_q/4)^2
  }{
  \omega_0^2
  }\right)^{1/3}.
\end{equation}

Evaluating the optimal coupling rate $g$ from Eq.~(\ref{eq:optg}) as a function of the number of storage modes $N$ numerically for $\omega_0/2\pi=4\,\mathrm{GHz}$, $\Gamma_q=1/(50\,\mu\mathrm{s})$ and $A=1$, we get the plot shown in Fig.~\ref{fig:VQplots}(a). For the mechanical resonator case, we use a realistic estimate of the mechanical quality factor $Q_{\mathrm{mech}}=\omega_0/\Gamma_r=10^9$ while for the microwave case, we assume that the quality factor of the resonators is comparable with that of the qubit, that is $\Gamma_r\approx \Gamma_q$. 

The fact that the optimal coupling can be reached in the electromechanical system is not obvious and is discussed in more detail in Sec.~\ref{sec:coupling}. The corresponding errors for both system, as given by Eq.~(\ref{eq:optError}) are plotted in Fig.~\ref{fig:VQplots}(b).

Using Eq.~(\ref{eq:VQ}), we can now calculate the quantum volume. Since the achievable circuit depth $d(N) = 1/N\varepsilon(N)$ plotted in Fig.~\ref{fig:VQplots}(c) is a decreasing function of $N$, the expression $\min(N,d(N))$ increases as long as $N<d(N)$, after which it starts to decrease. For simplicity, we relax the requirement that $N$ be integer, which allows us to approximately find the maximum of $\min(N,d(N))$ as the point where $N=d(N)$. For the specific parameters used above, this is shown graphically in Fig.~\ref{fig:VQplots}(c). The quantum volume of the electromechanical system is maximized for approximately $N=15$ resonators and reaches a value $V_Q = 220$. This is roughly three times higher than in the purely microwave on-chip system whose quantum volume reaches its maximum for about $N=8$ resonators.

More generally, we can estimate the quantum volume by solving the equation $N = d(N)$ while assuming $\Gamma_q\gg N\Gamma_r$ for the electromechanical system and $\Gamma_q\ll N\Gamma_r = N\Gamma_q$ for the microwave one. We then get
\[
 V_Q = \left\{
 \begin{array}{l}
 \left(\frac{2Q_q}{9\pi\sqrt{3A}}\right)^{1/2}\text{for the mechanical system,}\\
 \left(\frac{Q_q}{6\pi\sqrt{3A}}\right)^{2/5}\text{for the microwave system,}
 \end{array}
 \right.
\]
where $Q_q = \omega_0/\Gamma_q$ is the quality factor of the qubit, in our numerical estimate $Q_q = 1.25\times 10^6$. As alluded to before, we observe that this result scales quite weakly with the dimensionless constant $A$ which hides the details of the cross-talk error estimate.

\subsection{Additional error sources}

We should note that the estimates above neglect two other potential sources of error: relaxation of the qubit due to piezoelectric coupling to phonons in the substrate and relaxation of the resonators due to off-resonant coupling to the qubit (Purcell decay). We can expect the first effect to be negligible as long as the qubit frequency is within the phononic band gap because in that case it is protected against phonon radiation in the same way as the mechanical resonators. Finite element simulations confirm this intuition and show that the limit on the qubit coherence time from mechanical relaxation is above $100\,\mu\mathrm{s}$ for realistic phononic crystal designs, though this remains to be experimentally demonstrated. A conservative estimate of the Purcell decay contribution to the error probability can be obtained by approximating the excess relaxation rate in the resonators as $\Delta\Gamma_r = \Gamma_q (g/\delta)^2 = 4\Gamma_q N^2 g^2/\omega_0^2$. The corresponding increase in the effective gate error is $\Delta\varepsilon = 2\Delta\Gamma_r N T_s = \pi \Delta\Gamma_r N/g$. We need to compare this with the overall error probability. Using Eqs.~(\ref{eq:optg}) and (\ref{eq:optError}), we get
\[
  \frac{\Delta\varepsilon}{\varepsilon} 
  <
  \frac{2}{3}
  \left(\frac{
  \pi^2 N^5
  }{
  12A^2 Q_q^2
  }\right)^{1/3}.
\]
As shown above, the number of resonators maximizing the quantum volume is approximately $N = (2Q_q/9\pi\sqrt{3A})^{1/4}$ and therefore
\[
  \frac{\Delta\varepsilon}{\varepsilon}  <
  \frac{1}{9A}
  \left(\frac{
  8\pi\sqrt{A/3}
  }{
  3 Q_q
  }\right)^{1/4} \ll 1.
\]
This confirms that the Purcell decay effect is negligible for our purposes.

\section{Coupling of nanomechanical resonators to superconducting circuits}\label{sec:coupling}

\begin{figure*}
\begin{center}
\includegraphics[width=17.5cm]{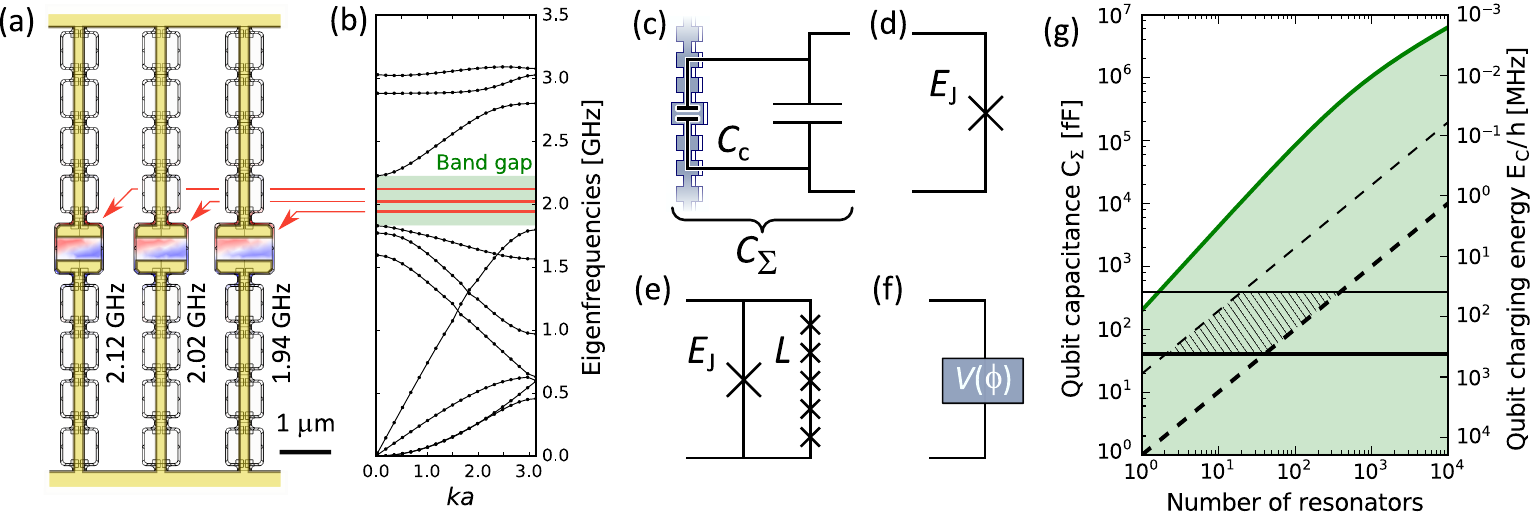}
\end{center}
\caption{(a) Drawing of several phononic crystal resonators and their coupling capacitors (in yellow). The red and blue color scheme represents the electrical potential generated by the localized mechanical modes. The lithographically defined variation in the dimensions of the resonator cell leads to different resonant frequencies, all of which lie within the band gap (b) of the surrounding phononic crystal. (c) Diagram representing the superconducting circuit's capacitive coupling to a phononic cavity resonator via the capacitor $C_c$. This is effectively connected in parallel with the rest of the circuit's capacitance, adding up to a total of $C_{\Sigma}$. The inductive part of the circuit may consist for example of a single Josephson junction (d), giving rise to a transmon qubit \cite{Koch2007} or a junction shunted by a nearly linear junction array (e), resulting in a fluxonium circuit \cite{Manucharyan2009}. In general, we consider the remainder of the circuit to be an arbitrary element (f) whose energy is diagonal in the $\phi$ eigenbasis. (g) The qubit capacitance values $C_{\Sigma}$ for which the optimal coupling from Eq.~(\ref{eq:optg}) can be reached, as a function of the number of resonators $N$, are indicated by the green region. The corresponding charging energies $E_C = e^2/2C_{\Sigma}$ are shown on the right axis. The thick black lines represent lower bounds on $C_{\Sigma}$ while the thin ones are upper bounds. The thick solid line indicates $C_{\Sigma} = 40\,\mathrm{fF}$, an approximate value above which a qubit with a frequency $\omega_0/2\pi=4\,\mathrm{GHz}$ can be considered to be in the transmon regime. The thick dashed line shows the limit due to the capacitance of the couplers, assuming a value of $C_1=1\,\mathrm{fF}$ per coupler. The thin solid line is an upper bound if the anharmonicity of the qubit is to be at least $50\,\mathrm{MHz}$. The thin dashed line represents the condition for the anharmonicity to exceed half of the nearest-neighbor detuning between the resonators. The shaded area indicates points consistent with all {the constraints above}.}
\label{fig:couplingDiagram}
\end{figure*}

We consider each of the mechanical resonators to be realized as a defect in the band gap of a one-dimensional phononic crystal, fabricated out of thin-film lithium niobate on silicon and coupled to two metal electrodes \cite{Arrangoiz-Arriola2016,ArrangoizArriola2018}, as shown in Fig.~\ref{fig:couplingDiagram}(a). Multiple resonators can be connected to the same qubit and separated in frequency space by choosing appropriate dimensions of the defect site in the phononic crystal. The resonance frequencies are chosen to lie within the gap in the crystal's band diagram plotted in Fig.~\ref{fig:couplingDiagram}(b). The electric field induced by the voltage between the two electrodes couples to the motion of the resonator via the piezoelectric effect in the lithium niobate film. This coupling is bilinear in the voltage $V$ and the effective displacement $x$ of the resonator mode, described by a Hamiltonian of the form $\op{H}_{\mathrm{int}}\propto\op{V}\op{x}$. Writing the displacement in terms of the mode's ladder operators $\op{a}$ and $\op{a}^{\dagger}$, we can express the Hamiltonian as
\[
  \op{H}_{\mathrm{int}} = q_{\mathrm{eff}} \op{V} (\op{a}+\op{a}^{\dagger}),
\]
where $q_{\mathrm{eff}}$ is a coupling parameter with a unit of charge. As part of the qubit's superconducting circuit, we can imagine the electrodes as a capacitor with a small capacitance $C_c$ connected in parallel with the rest of the circuit, as shown in Fig.~\ref{fig:couplingDiagram}(c-f), giving rise to a total effective capacitance $C_{\Sigma}$. In this picture, we explicitly consider only a single mechanical resonator but we can imagine that the remaining $N-1$, each connected to its own set of coupling electrodes, simply contribute to the total capacitance $C_{\Sigma}$. Only a single resonator is resonant with the qubit at any given time and the presence of the $N-1$ off-resonant ones does not change the following analysis. We should also note that since each resonator is suspended and only connected to the bulk of the chip by the phononic crystal with a wide band gap, there is for all practical purposes no direct phononic coupling between the resonators. This high degree of isolation is enabled by the wide phononic bandgap and the complete absence of phonon propagation in vacuum, something which does not have a parallel in microwave systems.

The qubit itself may be one of several types of superconducting devices, for example a transmon \cite{Koch2007} or a fluxonium \cite{Manucharyan2009}. In both of these, the capacitance $C_{\Sigma}$ is shunted by a non-linear inductive component. In the case of the transmon, this is a single Josephson junction (see Fig.~\ref{fig:couplingDiagram}(d)), whereas for the fluxonium, it is a junction in parallel with a linear inductor (see Fig.~\ref{fig:couplingDiagram}(e)). In both cases, the circuit can be seen generally as a capacitance $C_{\Sigma}$ connected in parallel to a component with an energy $V(\phi)$ that depends only on the phase variable $\phi$ (see Fig.~\ref{fig:couplingDiagram}(f)).

The strength of the coupling between the qubit and the mechanical mode can be characterized by the matrix element $\hbar g = \langle g 1| \op{H}_{\mathrm{int}} |e 0\rangle$, where $|g1\rangle = |g\rangle\otimes|1\rangle$ is the tensor product of the qubit's ground state $|g\rangle$ with the phononic single-photon Fock state $|1\rangle$. Similarly, $|e0\rangle = |e\rangle\otimes|0\rangle$ is a combination of the qubit's first excited state $|e\rangle$ and the phononic vacuum state $|0\rangle$. The coupling strength parameter $g$ is then given by
\[
  g = \frac{2e q_{\mathrm{eff}}}{\hbar C_{\Sigma}} \langle g|\op{n}|e\rangle,
\]
where $\op{n}$ is the Cooper-pair number operator. As we will now show, there is an upper limit on $g$ which depends only on $C_{\Sigma}$ and the qubit frequency $\omega_0$, independently of the exact nature of the circuit's inductive part. This limit follows from the well-known \emph{Thomas-Reiche-Kuhn sum rule} \cite{Sakurai1995MQM} but we show its derivation here for completeness. We start with the circuit's Hamiltonian
\[
  \op{H}=4E_C\op{n}+V(\op{\phi}),
\]
where $E_C = e^2/2C_{\Sigma}$ is the charging energy of the qubit. We then observe that due to the identity $\exp(\mathrm{i}u\op{\phi})\op{n}\exp(-\mathrm{i}u\op{\phi})=\op{n}+u$, the ground state energy $E_0$ of the modified Hamiltonian $\op{H}(u)=4E_C(\op{n}+u)^2+V(\op{\phi})$ does not depend on $u$ \footnote{This argument fails if the variable $\phi$ is cyclic because then the unitary we applied to the Hamiltonian does not preserve the periodic boundary conditions and is therefore not allowed. But in such a case the dependence of the eigenenergies on $n_g$ means charge sensitivity of the qubit. This is typically avoided \emph{by design} so we will assume that the dependence is negligible in any case.}. In particular, the second derivative of $E_0$ with respect to $u$ at $u=0$ is then zero. Expressing this derivative using perturbation theory, we get
\[
  8E_C - 2\sum_{i>0}\frac{|\langle\varphi_i|8E_C\op{n}|\varphi_0\rangle|^2}{E_i-E_0} = 0,
\]
where $|\varphi_i\rangle$ are the eigenstates of the circuit Hamiltonian $\op{H}$ and $E_i$ their eigenenergies. Specifically, $|\varphi_0\rangle=|g\rangle$, $|\varphi_1\rangle=|e\rangle$ and $E_1-E_0=\hbar\omega_0$. All the terms in the sum are non-negative and therefore
\[
  8E_C - 2\frac{|\langle\varphi_1|8E_C\op{n}|\varphi_0\rangle|^2}{E_1-E_0} \ge 0.
\]
From here it follows that
\[
  g \le q_{\mathrm{eff}}\sqrt{\frac{\omega_0}{2\hbar C_{\Sigma}}}.
\]
The maximum is reached for a purely linear circuit, that is, one with $V(\op{\phi}) = E_L\op{\phi}^2/2$. We will consider the case where the circuit is a weakly non-linear transmon qubit which can closely approach the theoretical limit above. While it is possible for a strongly non-linear circuit with a lower $C_{\Sigma}$ to reach a stronger coupling than a transmon with a higher $C_{\Sigma}$, we will see that the required coupling is compatible with a weakly non-linear system and it is therefore sufficient to consider a transmon.

To achieve the optimal coupling rate given by Eq.~(\ref{eq:optg}), the capacitance of the qubit needs to be {at most}
\[
  C_{\Sigma} = \frac{2q_{\mathrm{eff}}^2}{\hbar\omega}
  \left(
  \frac{
  4 A N^2\omega_0
  }{
  \pi(N\Gamma_r+3\Gamma_q/4)
  }\right)^{2/3}.
\]
To estimate the parameter $q_{\mathrm{eff}}$, we consider the coupling geometry analyzed in \cite{Arrangoiz-Arriola2016} (see Fig.~\ref{fig:couplingDiagram}(a)). There the total capacitance of the simulated ciruit was $C_{\Sigma} = 200\,\mathrm{fF}$ and the calculated coupling roughly $g/2\pi=10\,\mathrm{MHz}$ at a frequency $\omega_0/2\pi = 2\,\mathrm{GHz}$. As the weakly non-linear circuit effectively reaches the upper bound on $g$ derived above, we can calculate $q_{\mathrm{eff}}=\sqrt{2\hbar C_{\Sigma} g^2/\omega_0} = 4\times 10^{-21}\,\mathrm{C}$. We expect that this number could be made higher by further optimization of the coupler geometry and it can certainly be made smaller if desired, so this value is a conservative upper bound on $q_{\mathrm{eff}}$. 

In practice, we cannot make $C_{\Sigma}$ arbitrarily low, mainly due to the two following constraints: Each of the $N$ qubit-resonator couplers has an associated capacitance $C_1$ and $C_{\Sigma}$ therefore has to be at least $N C_1$. Finite element simulations indicate that for the coupler design shown in Fig.~\ref{fig:couplingDiagram}(a), $C_1$ is on the order of $1\,\mathrm{fF}$. Furthermore, $C_{\Sigma}$ needs to be high enough to bring the qubit into the transmon regime where $E_J\gg E_C$. Since $2E_J/E_C = (\hbar\omega_0 C_{\Sigma}/e^2)^2$, the minimal $C_{\Sigma}$ required to achieve $E_J/E_C\gg 1$ is approximately $C_{\Sigma}=4e^2/\hbar\omega_0$. Assuming $\omega_0/2\pi = 4\,\mathrm{GHz}$, this equals approximately $40\,\mathrm{fF}$. 

We plot the maximum capacitance $C_{\Sigma}$ consistent with the optimal coupling $g$ in Fig.~\ref{fig:couplingDiagram}(g) and observe that even for low numbers of resonators, it lies above the two aforementioned lower limits. 

Finally, the transmon circuit also needs to have a sufficiently large anharmonicity to be useful as a qubit and to satisfy the assumption we made when deriving the cross-talk error estimate in Sec.~\ref{sec:gateError}. Namely that the spurious transition to the second excited state can be kept detuned by at least half of the nearest-neighbor detuning between the resonators $\omega_0/2N$. As the anharmonicity in a weakly non-linear circuit is approximately given by the charging energy $E_C=e^2/2C_{\Sigma}$, this gives upper limits on $C_{\Sigma}$ which are shown by the thin solid and dashed lines in Fig.~\ref{fig:couplingDiagram}(g).

The three upper limits (one set by the necessary coupling $g$, the other two by the minimal anharmonicity) and two lower limits (due to the transmon condition and the capacitance of the couplers) define the shaded region in Fig.~\ref{fig:couplingDiagram}(g) which extends all the way to approximately $N = 200$. Hence, even taking the practical constraints discussed here into account, we expect that the optimal coupling given by Eq.~(\ref{eq:optg}) can be achieved in devices with a significant number of storage modes.

\section{Conclusions}
We have discussed a hybrid quantum information processing architecture which combines a superconducting qubit acting as a processor and multiple nanomechanical resonators based on phononic crystal cavities for information storage, coupled directly to the qubit. The phononic crystal resonators are uniquely suited to make the storage modes both very long lived and compact. This, together with the fact that only the processing qubit needs to be externally controlled, is beneficial for scaling.

We have carefully analyzed the trade-off between two major sources of error in such a system -- gate cross-talk and decoherence -- and found an optimal value for the qubit-resonator coupling which minimizes the estimated error. We then showed that the calculated optimal coupling can be reached even if practical constraints on the system are taken into account. To analyze the performance of the proposed system in quantum computing applications, we estimated its quantum volume and found it to be around $220$. That is, a system of this kind with approximately $15$ stored qubits could run a quantum circuit with a depth of $15$ before the error probabilities become significant. For comparison, we also analyzed an analogous system using on-chip microwave resonators for information storage and found that its quantum volume is smaller by about a factor of $3$ due to the lower quality factors of the resonators.

We emphasize that the results derived here for the electromechanical system apply equally to any other implementation where the storage modes are significantly longer-lived than the qubit. For instance, storage in high-quality 3d microwave cavities could in principle achieve the same performance. However, thanks to the very small size of the mechanical resonators, our proposed approach does not suffer from the scaling difficulties which may arise in a system with a large number of 3d cavities.

{The analysis presented here considers the platform in a rather implicit way. Much work is required to understand what error correcting codes and algorithms are best suited for this type of hardware. This will be the focus of future studies.}

Finally, we note that though we've analyzed this electromechanical platform mainly in the context of quantum {information processing}, we expect such long-lived compact quantum memories to find applications in quantum repeater systems \cite{Briegel1998,Muralidharan2016}. In this context where storage is the primary purpose of the device and not merely a necessity enforced by the sequential gate execution, using mechanical modes with a high quality factor achieves a true advantage over on-chip microwave circuits. In the simplest quantum repeater schemes operating without error correction \cite{Muralidharan2016}, the memory needs to hold information for an extended period of time until entangled qubit pairs are successfully distributed over all sections of the long quantum link. Due to transmission losses, the entanglement distribution scheme is non-deterministic and needs to be heralded. For long distances, the average time until success may be significantly longer than the propagation time over the whole link, necessitating very high quality quantum memories.

\begin{acknowledgments}
{The authors thank Patrick Hayden and Alex Wollack for insightful discussions.  This work is supported by the U.S.
government through the Office of Naval Research under
MURI No. N00014-151-2761, and the National Science
Foundation under grant No. ECCS-1708734. This work was supported by the David and Lucille Packard Fellowship. M.P. is partly supported by the Swiss National Science Foundation. P.A.A. is partly supported by the Stanford Graduate Fellowship.}
\end{acknowledgments}

\end{document}